# Potential-Induced Dynamic Coordination of Nonmetal Atoms Directly Bound to Metal Centers in Graphene-Embedded Single-Atom Catalysts and Its Implications

Jiahang Li[1], Suhang Li[1], Chong Yan[1], Qinzhuang Liu[1], Jiajun Yu[1*], Dongwei Ma[1*]
[1]Anhui Provincial Collaborative Innovation Center for Advanced Functional Composite Materials, *College of Physics and Electronic Engineering, Huaibei Normal University, Huaibei 235000, China*

## ABTRACT

Electrode-potential-induced dynamic coordination is an essential factor governing the performance of graphene-embedded single-atom catalysts (SACs). While previous studies have primarily centered on structural dynamics at the metal site, the response of its coordinated nonmetal atoms remains largely unexplored. Here, using Ni SACs with mixed nitrogen/carbon coordination ($NiN_{4-x}C_x$) as representatives, we investigate potential-driven hydrogenation of metal-center-coordinated nonmetallic atoms through constant-potential density functional theory and *ab initio* molecular dynamics. We find that the C sites directly bound to Ni undergo potential-driven hydrogenation, whereas hydrogenation at N sites is thermodynamically unfavorable. Taking $NiNC_3$ as a representative system, we demonstrate that these hydrogenation processes proceed with accessible kinetic barriers and obey the Brønsted-Evans-Polanyi relation. The resulting dynamic coordination reshapes the Ni 3d and $d_{z^2}$ orbitals, modulates the stability of the active center, and weakens molecular adsorption through combined electronic and steric effects. These findings reveal that electrode potential and solvent not only regulate the metal center but also dynamically reconfigure its coordination environment, offering novel mechanistic insights into potential-induced coordination dynamics and guiding the rational design of coordination-engineered SACs.

*Corresponding author. E-mail: yujiajun@chnu.edu.cn (J. Yu)
*Corresponding author. E-mail: madw@chnu.edu.cn, dwmachina@126.com (D. Ma)

## I. INTRODUCTION

Single-atom catalysts (SACs) have emerged as a prominent class of atomically dispersed heterogeneous catalysts [1,2]. Among them, graphene-embedded SACs have attracted particular attention owing to their well-defined active sites, excellent electrical conductivity, large surface area, and tunable coordination environments [3,4]. Consequently, these catalysts have been extensively investigated for a wide range of catalytic reactions [5-7]. In particular, coordination engineering—the systematic tuning of the type, number, or spatial arrangement of nonmetallic atoms around the metal center—has proven to be an effective strategy for optimizing catalytic activity and selectivity. By modulating the metal electronic structure, such ligand-field control enables fine adjustment of adsorption energies, reaction pathways, and overall stability [8-12].

Recent studies have revealed that, under realistic operating conditions, the local structure of single-atom catalytic sites is not static but dynamically adapts to external factors such as temperature and electrode potential, particularly in graphene-embedded SACs [12-17]. In electrochemical environments, variations in electrode potential inject or withdraw electrons from both the graphene support and the metal center, thereby modulating their oxidation states and the surface charge of the graphene [18-20]. This electronic modulation can strengthen or weaken the interactions between the metal center and its coordinated atoms in the graphene lattice, and can further drive the adsorption or desorption of ligands, such as $H_2O$, $OH^-$, $H^+$, and reaction intermediates, at the metal site [21-26]. Consequently, the coordination sphere of the SAC, including both the number and identity of neighboring atoms, evolves with potential, profoundly altering the electronic structure of the active site and, in turn, its catalytic activity and selectivity. A representative example is the Fe–$N_4$ SAC, where a single Fe atom is coordinated to four pyridinic nitrogen atoms embedded in graphene. Under oxygen reduction reaction conditions, the actual active site is proposed to be an Fe atom axially coordinated with $H_2O$ [27,28], OH, or O [29,30] species generated by water adsorption or oxidation at positive potentials, depending on the applied potential and pH.

Despite extensive experimental and theoretical efforts to elucidate electrode-potential-induced dynamic coordination in graphene-embedded SACs, most studies have focused primarily on structural changes at the metal center itself [12,14,31]. Whether the applied potential can also trigger re-coordination of the heteroatoms that anchor the metal center to the graphene support—and how such transformations influence the properties of the metal site—remains largely unexplored. This gap presents a critical challenge to achieving a comprehensive understanding of potential effects and to advancing coordination engineering in SACs.

Here, using representative $NiN_{4-x}C_x$ SACs [32-34], we investigate the dynamic coordination of nonmetal atoms directly bound to metal centers in graphene-embedded SACs under the influence of electrode potential and interfacial water, based on constant-potential density functional theory (CP-DFT) and *ab initio* molecular dynamics (AIMD) simulations with explicit water. We first examine the thermodynamics and kinetics of hydrogenation at both C and N atoms directly bonded to the Ni active center, revealing that C sites can be dynamically hydrogenated by the applied potential in the presence of interfacial $H_2O$. Using $NiNC_3$ as a model system, we further explore the stability, electronic structure, and chemical activity of the Ni center. It is demonstrated that potential-induced dynamic hydrogenation of the coordinating C atoms substantially modifies the stability, electronic configuration, and reactivity of the Ni active site, thereby affecting catalytic performance. This phenomenon is expected to be general among graphene-embedded atomically dispersed metal catalysts—including single-atom, dual-atom, and small-cluster systems—underscoring the importance of accounting for the dynamic coordination of nonmetallic atoms bound to metal centers to achieve a comprehensive understanding of potential effects and to guide coordination engineering.

## II. COMPUTATIONAL METHODS

Spin-polarized DFT calculations were performed using the Vienna *ab initio* Simulation Package (VASP) [35]. The projector augmented wave (PAW) method was employed to describe the interaction between valence electrons [36] and ionic cores, and the Perdew-Burke-Ernzerhof (PBE) functional was used for the exchange-correlation energy [37]. A plane-wave cutoff energy of 400 eV was applied. Long-range van der Waals interactions were treated with the DFT-D3 correction [38]. Geometry optimizations were converged to $10^{-5}$ eV in total energy and 0.02 eV $Å^{-1}$ in forces. The $NiN_{4-x}C_x$ SACs model is constructed based on a 5×3 rectangular graphene supercell, with a Monkhorst-Pack *k*-point mesh of 3×3×1 for structural optimization and 6×6×1 for density of states (DOS) calculations [39]. A vacuum layer of at least 16 Å was applied to eliminate interactions between periodic images. The crystal orbital Hamilton population (COHP) analysis was carried out using the LOBSTER package, and the integrated COHP (ICOHP) values were used to quantify the strength and orbital contributions of individual chemical bonds [40]. The VASP output files were post-processed and analyzed with VASPKIT [41].

The effect of electrode potential was incorporated by optimizing both the atomic coordinates and the electron count at the target potential relative to the standard hydrogen electrode (SHE) [18,42]. The electrode potential of the charged slab referenced to the standard hydrogen electrode (U vs. SHE) was calculated as:

$$U_{SHE} = -4.6 - \phi/e$$

where –4.6 is the work function of the SHE in VASPsol [43,44], and $-\phi$ denotes the work function of the charged system.

The relationship between the potentials versus SHE and the reversible hydrogen electrode (RHE) was given by:

$$eU_{RHE} = eU_{SHE} + k_BT \times pH \times \ln 10$$

Herein, the pH was set to 6.8 in the computational procedures, representing typical conditions for electroreduction reactions such as $CO_2$, NO, and $NO_3^-$ reduction [45,46].

The H adsorption free energy ($\Delta G_{*H}$) has been calculated according to the computational hydrogen electrode (CHE) model [47]:

$$\Delta G = \Delta E_{*H} + \Delta E_{ZPE} - T\Delta S + eU_{RHE}$$

where $\Delta E_{*H}$ is the adsorption energy obtained from CP-DFT total energies, $\Delta E_{ZPE}$ is the zero-point energy correction, and $T\Delta S$ (with $T$ = 298.15 K) is the entropic contribution. $\Delta E_{ZPE}$ and $T\Delta S$ were determined from vibrational-frequency calculations for the adsorbed intermediates and from NIST database values for the gas-phase molecules [48].

The energy barrier for water dissociation was evaluated using AIMD simulations with the slow-growth approach [49], sampling only the Γ point of the Brillouin zone without symmetry constraints. Collective variables were defined by the distance changes of atoms involved in the elementary reaction, and a transformation velocity of 0.0008 Å was applied. The simulations employed a constant-potential hybrid-solvation (CP-HS) model [50,51], as described in Fig. S1 of the Supplemental Material [52]. To enhance numerical stability, the hydrogen mass was set to 2 amu, and a 1 fs time step was used. The system temperature was maintained at 300 K with a Nosé-Hoover thermostat [53].

## III. RESULTS AND DISCUSSION

Various $NiN_{4-x}C_x$ systems have been experimentally synthesized and adopted as the model in this study [32-34], with their atomic structure shown in Fig. 1(a). We investigated the electrode-potential-induced dynamic coordination of Ni-bound carbon atoms in aqueous media using AIMD simulation. Six layers of explicit $H_2O$ molecules were placed above the graphene substrate to capture realistic solvation effects. AIMD provides a detailed description of interfacial solvation and surface chemical bonding. The atomic density profile in Fig. 1(b) reveals an increased density in the interfacial water layer, attributable to slower diffusion of species at the interface, consistent with previous reports [27]. The water layer interacts only weakly with the graphene substrate, maintaining an interfacial distance of about 3 Å, indicative of a hydrophobic surface character [54,55].

We next evaluated the thermodynamics of hydrogenation at both C and N sites of $NiN_{4-x}C_x$ systems under various electrode potentials, using the free energy change $\Delta G_{*H}$ defined in the *Computational Methods* section. As shown in Fig. 2(a), the hydrogenation at the N site is thermodynamically highly unfavorable at 0 $V_{RHE}$ for all systems; in particular, $NiN_4$, $NiN_3C$, $NiN_2C_2$-o1, and $NiN_2C_2$-o2 exhibit $\Delta G_{*H}$ values of about 1 eV. Lowering the electrode potential (making it more negative) can render the N-site hydrogenation more favorable, with $NiNC_3$ showing the most pronounced effect, reaching a $\Delta G_{*H}$ of −0.22 eV at −1.2 $V_{RHE}$. Nevertheless, as discussed below, the N-site hydrogenation on $NiNC_3$ at −1.2 $V_{RHE}$ is still unlikely to occur because of kinetic limitations.

In contrast, hydrogenation at the C site is far more thermodynamically favorable than at the N site. Taking $NiNC_3$ as an example, whose atomic configurations are shown in Fig. 2(b), the $\Delta G_{*H}$ for the N site is 0.69 eV at 0 $V_{RHE}$, whereas the $\Delta G_{*H}$ values for the $C_1$, $C_2$, and $C_3$ sites are −0.48, −0.02, and −0.19 eV, respectively. This also indicates that, at 0 $V_{RHE}$, the $C_1$ site is the most favorable for hydrogenation, being markedly more exothermic than the other three atoms directly bound to the Ni center. Moreover, as the electrode potential decreases, the $\Delta G_{*H}$ of the $C_1$ site becomes increasingly negative, reaching −1.20 eV at −1.2 $V_{RHE}$. The $C_2$ and $C_3$ sites follow the same trend, though their adsorption strengths remain weaker than that of $C_1$ across all potentials.

Building on these results, we further examined the second hydrogenation step, starting from the configuration with the first hydrogen atom adsorbed at the $C_1$ site ($NC_3$-$1H_{C1}$, Fig. 2(b)). For this second hydrogenation, the $C_3$ site is consistently more favorable than the $C_2$ site at all electrode potentials. Hydrogenation becomes thermodynamically feasible when the electrode potential reaches about −0.8 $V_{RHE}$, where the $\Delta G_{*H}$ values for the $C_2$ and $C_3$ sites are −0.40 and −0.64 eV, respectively. Interestingly, at −1.2 $V_{RHE}$, the hydrogenation of a third H atom at the $C_3$ site has a $\Delta G_{*H}$ of −0.49 eV, indicating that under these conditions all three C atoms can be hydrogenated. These findings suggest that, under sufficiently negative electrode potentials, the C atoms directly coordinated to the Ni site in $NiNC_3$ can be dynamically hydrogenated by protons from the aqueous solvent. Similar behavior is expected for other $NiN_{4-x}C_x$ systems, as supported by the $\Delta G_{*H}$ values summarized in Fig. 2(a); the corresponding atomic configurations are provided in Fig. S2 of the Supplemental Material [52].

We then examined the mechanism underlying the potential-induced stronger adsorption of H at the C site compared with the N site. To this end, we computed the adsorption energy $\Delta E_{*H}$ from CP-DFT total energies, which reflects the interaction strength between the H atom and the charged $NiNC_3$. As shown in Fig. 2(c), taking the $C_1$ site as an example, $\Delta E_{*H}$ is ~1 eV more negative for $C_1$ than for the N site at each electrode potential, indicating substantially stronger H−C binding. Another notable trend is that, as the electrode potential becomes more negative, $\Delta E_{*H}$ shifts toward less negative values, implying a weakened interaction between H and the charged $NiNC_3$ at increasingly reducing potentials. To rationalize these observations, we calculated the projected DOS (PDOS) of the p states for both $C_1$ and N. Fig. 2(d) shows that the p-band center (relative to the Fermi level) of N lies much lower than that of $C_1$: the $C_1$ p-band center ranges from −2.9 to −3.6 eV, while the N p-band center ranges from −5.2 to −5.8 eV. Moreover, lowering the potential shifts the p-band centers of both $C_1$ and N downward relative to the Fermi level. These results are consistent with the p-band-center theory, which predicts that a higher (less negative) p-band center correlates with stronger adsorbate binding [56-58]. A similar shift of the p-band center with varying electrode potential is also observed for the $C_2$ and $C_3$ atoms, as shown in Fig. S3 of the Supplemental Material [52].

The origin of the dynamic H coordination at the Ni-coordinated C site was further examined. Undoubtedly, the H atom originates from the hydrolysis of $H_2O$ in aqueous solution. To probe this, we investigated the kinetic barrier for the hydrolysis of $H_2O$ leading to the formation of H* and $OH^-$ using the slow-growth simulation. To establish a representative solvation structure around the catalytic site, an AIMD simulation was first carried out prior to the slow-growth process to ensure proper equilibration of both energy and temperature (see Fig. S4 in the Supplemental Material [52]) In addition, taking the hydrogenation of the C1 site as example, the stretching of the O−H bond in a neighboring $H_2O$ molecule and the shortening of the C−H bond was chosen as collective variables (see Fig. S5 in the Supplemental Material [52]). The slow-growth simulation was terminated once the C−H bond length approaches that of the statically optimized structure, as shown in Fig. S6 of the Supplemental Material [52].

The slow-growth results for $NiNC_3$ systems are summarized in Fig. 3, Fig. S7, and Table S1 in the Supplemental Material [52]. At 0 $V_{RHE}$, the kinetic barrier for hydrogenation at the $C_1$ site is 0.78 eV (Fig. 3(a)), indicating that this process is kinetically feasible at room temperature. However, as discussed below, the second hydrogenation is inaccessible under these conditions due to kinetic limitations. At −0.4 $V_{RHE}$, the kinetic barrier at the $C_1$ site decreases to 0.59 eV (Fig. 3(b)), consistent with the enhanced thermodynamics shown in Fig. 2(a). For the second hydrogenation, we evaluated the $C_3$ site, which is more thermodynamically favorable than $C_2$. Nevertheless, the barrier for this step rises to ~1.2 eV (Fig. S7(a)), suggesting that simultaneous hydrogenation of both $C_1$ and $C_3$ is unlikely at room temperature under −0.4 $V_{RHE}$.

At −0.8 $V_{RHE}$, the barrier for $C_1$ hydrogenation further decreases to ~0.5 eV (Fig. S7(b)), and hydrogenation at $C_2$ and $C_3$ also becomes kinetically feasible, with barriers of 0.75 eV (Fig. S7(c)) and 0.55 eV (Fig. S7(d)), respectively. Following the first hydrogenation at $C_1$, the second hydrogenation at either $C_2$ or $C_3$ is also accessible, as shown in Fig. 3(c) and Fig. 3(d). In contrast, the third hydrogenation is strongly hindered, with a barrier of 1.39 eV (Fig. S7(e)). Finally, at −1.2 $V_{RHE}$, given that both the first and second hydrogenations are highly favorable thermodynamically (Fig. 2(a)), we focused on the third hydrogenation at the $C_2$ site. As shown in Fig. 3(e), this step proceeds with an energy barrier of 0.85 eV, indicating feasibility at room temperature.

Overall, Fig. 3(f) summarizes the relationship between the hydrogenation barrier and the hydrogenation free energy. A clear linear correlation is obtained ($E_a$ = 1.17$\Delta G_{*H}$ + 1.37), with coefficient of determination of 0.903, confirming the Brønsted-Evans-Polanyi (BEP) relation [59,60]. On the basis of this BEP relation, we conclude that the C site directly coordinated to the Ni center in $NiNC_3$ and other $NiN_{4-x}C_x$ systems undergoes dynamic H coordination driven by both electrode potential and solvent, which is expected to influence the stability, electronic structure, and the chemical and catalytic activities of the Ni active center.

In the following, we investigate the effect of dynamic hydrogenation at the C site on the electrochemical stability of the Ni center in $NiNC_3$. The leaching process, schematically illustrated in Fig. 4(a), can be expressed as:

$$\mathrm{NiN_{4-x}C_xH_y} \leftrightarrow \mathrm{N_{4-x}C_xH_y} + \mathrm{Ni^{2+}(aq)} + 2e^-$$

and the corresponding free energy change, i.e., the leaching energy, is given by [51] (for details see Note 1 of the Supplemental Material [52]):

$$\Delta G = G\left(\mathrm{N_{4-x}C_xH_y}\right) + G(\mathrm{Ni(s)}) + 2eU_0 - 2eU_{\mathrm{RHE}} - G\left(\mathrm{NiN_{4-x}C_xH_y}\right)$$

The calculated leaching energies for all hydrogenation configurations at the considered electrode potentials are summarized in Fig. 4(b). In all cases, the bare $NiNC_3$ exhibits the highest (most positive) leaching energies, making it thermodynamically least favorable for leaching. Hydrogenation at the C site markedly lowers the leaching energies, and the general trend is that the leaching energy decreases progressively with increasing numbers of H atoms. Notably, when all C atoms are hydrogenated, the leaching energies become negative. These results indicate that dynamic H coordination at the C site thermodynamically facilitates Ni dissolution under operating electrochemical conditions.

We further analyzed the orbital interactions between the Ni 3d states and the coordinated N 2p, C 2p, and H 1s orbitals at −0.8 $V_{RHE}$. As shown in Fig. 4(c) (top and middle panels), the strong orbital interaction near −1.2 eV observed in bare $NiNC_3$ disappears upon hydrogenation at the $C_1$ and $C_3$ sites. This trend is further confirmed by the COHP analysis in Fig. 4(d): in bare $NiNC_3$, the states around −1.2 eV correspond to pronounced bonding interactions, which vanish when $C_1$ and $C_3$ are hydrogenated. Consistently, the integrated COHP (ICOHP) shows that hydrogenation at $C_1$ and $C_3$ reduces the ICOHP by ~1.6 eV. In contrast, the Ni−H interaction contributes an ICOHP of −2.30 eV, indicating that the Ni bonding strength in the hydrogenated configuration surpasses that in bare $NiNC_3$ owing to the additional Ni−H interaction. Taken together, these results suggest that although dynamic H coordination thermodynamically destabilizes the Ni center, it can kinetically enhance its stability by strengthening local Ni−H bonding.

Dynamic H coordination at the C sites is expected to modify the electronic structure of the Ni active center. This effect is evaluated through shifts in the band centers of the Ni 3d and $d_{z^2}$ orbitals [61]. Fig. 5(a) displays the PDOS of Ni for bare $NiNC_3$ and its hydrogenated configurations under various electrode potentials. Both orbitals exhibit pronounced shifts upon hydrogenation, with the overall trend being a significant downshift of the states. For example, at −0.8 $V_{RHE}$, the d-band center decreases from −1.02 eV to −1.71 eV, while the $d_{z^2}$-band center shifts from −0.86 eV to −1.77 eV. A comparable downshift of both the d- and $d_{z^2}$-band centers is also observed at −0.4 and −1.2 $V_{RHE}$. This systematic downshift of the Ni 3d and $d_{z^2}$ orbitals can be attributed to the upshift of the Fermi level (Fig. S8 in the Supplemental Material [52]), resulting from electron donation from the adsorbed H atom to the $NiNC_3$ system. This electron transfer also renders the potential of zero charge more negative. Furthermore, hydrogenation of the C site makes the Ni 3d and $d_{z^2}$ orbitals more delocalized, as reflected by the broadening of the d and $d_{z^2}$ states with increasing numbers of hydrogenated C atoms.

Finally, we examined the effect of dynamic hydrogenation of the C site on the chemical activity of the Ni center, focusing on NO adsorption. As shown in Fig. 5(b), the adsorption energies of NO on hydrogenated $NiNC_3$ are substantially higher (more positive) than those on the bare $NiNC_3$, indicating that dynamic H coordination at the C site markedly modulates the chemical activity of the Ni center. Notably, this variation in chemical activity exhibits a certain degree of correlation with the d- or $dz^2$-band centers, where a lower band center generally indicates weaker adsorption strength. In addition, steric hindrance also plays an important role in the reduced NO adsorption. Specifically, as shown in Fig. S9 of the Supplemental Material [52], at −1.2 $V_{RHE}$ the Ni active center becomes sterically blocked, preventing the NO molecule from being captured. Since molecular adsorption underpins catalytic activity and selectivity, these results further suggest that dynamic hydrogenation of the coordinating atoms surrounding the metal center can regulate both activity and selectivity, warranting further investigation.

## IV. CONCLUTIONS

In summary, we have shown that electrode potential and interfacial water drive dynamic hydrogenation of carbon atoms coordinated to Ni in $NiN_{4-x}C_x$ systems, whereas nitrogen sites remain largely inactive. Using $NiNC_3$ as a representative case, the process follows a Brønsted-Evans-Polanyi relation with feasible barriers under reducing potentials. Dynamic hydrogenation decreases the thermodynamic stability of the Ni center by lowering leaching energies, while local Ni−H interactions provide partial kinetic stabilization. Simultaneously, the electronic structure is significantly altered, with pronounced downshifts of the Ni 3d- and $d_{z^2}$-band centers, and the chemical activity is modulated through combined electronic and steric effects. These findings demonstrate that electrode-potential-induced dynamic coordination extends beyond the metal atom to its anchoring environment, offering a new perspective for tailoring the stability and reactivity of graphene-embedded SACs. This concept is expected to be broadly applicable to other graphene-based atomically dispersed catalysts, including double-atom and cluster systems.

## CONFLICT OF INTEREST

There are no conflicts of interest to declare.

## ACKNOWLEDGMENTS

The work was supported with the Excellent Scientific Research and Innovation Team of the Education Department of Anhui Province (No. 2024AH010027). Middle-aged and young teachers' training action discipline (major) leader cultivation project (No. DTR2023022).

FIGURES

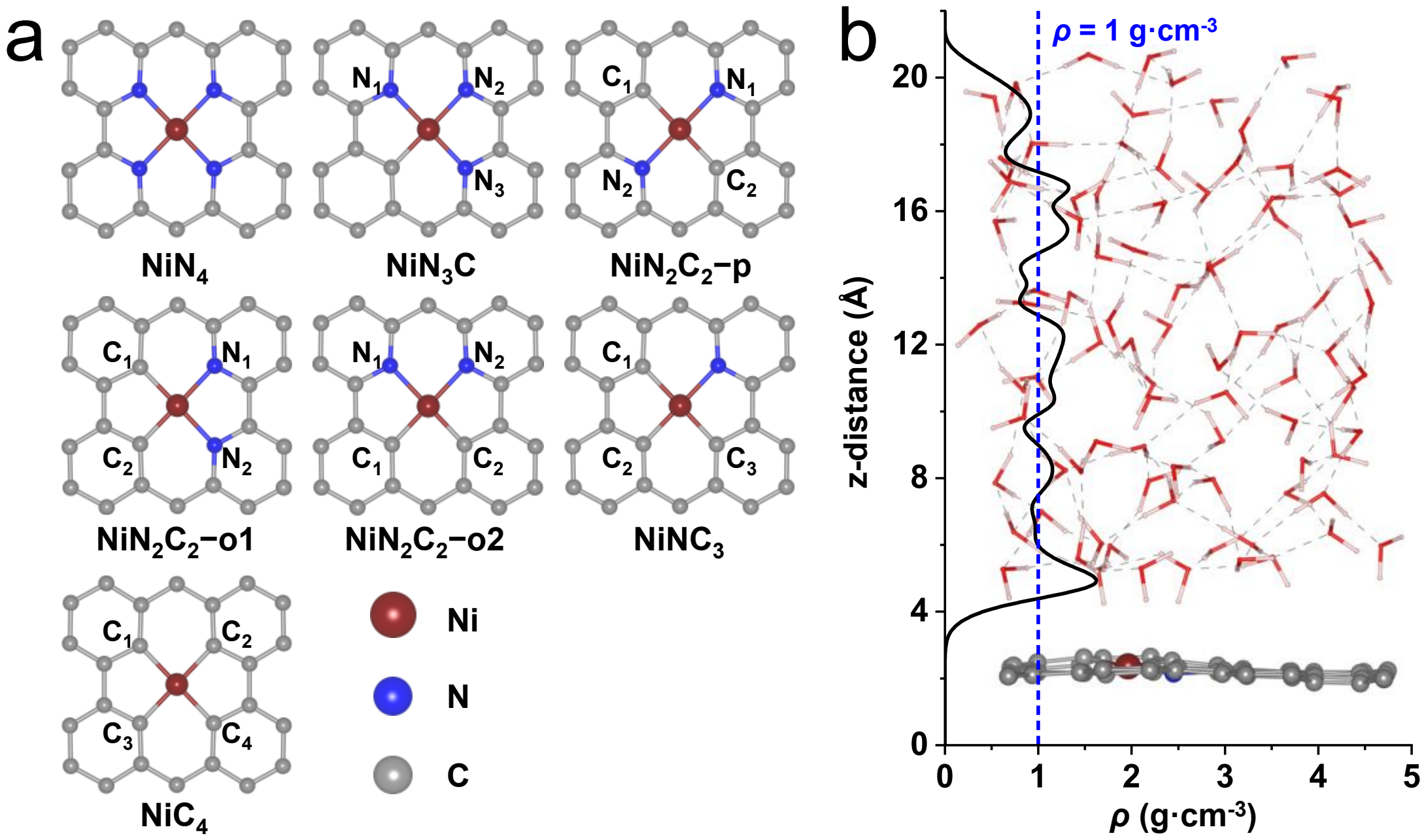

**Figure 1.** (a) Atomic models of Ni SACs with mixed nitrogen/carbon coordination ($NiN_{4-x}C_x$), including $NiN_4$, $NiN_3C$, $NiN_2C_2$-p, $NiN_2C_2$-o1, $NiN_2C_2$-o2, $NiNC_3$, and $NiC_4$. Ni, N, and C atoms are shown in vermilion, blue, and gray, respectively. The representative labeling of C and N sites is indicated. (b) Atomic density profile (ρ) of interfacial water molecules along the z-direction for $NiNC_3$ systems.

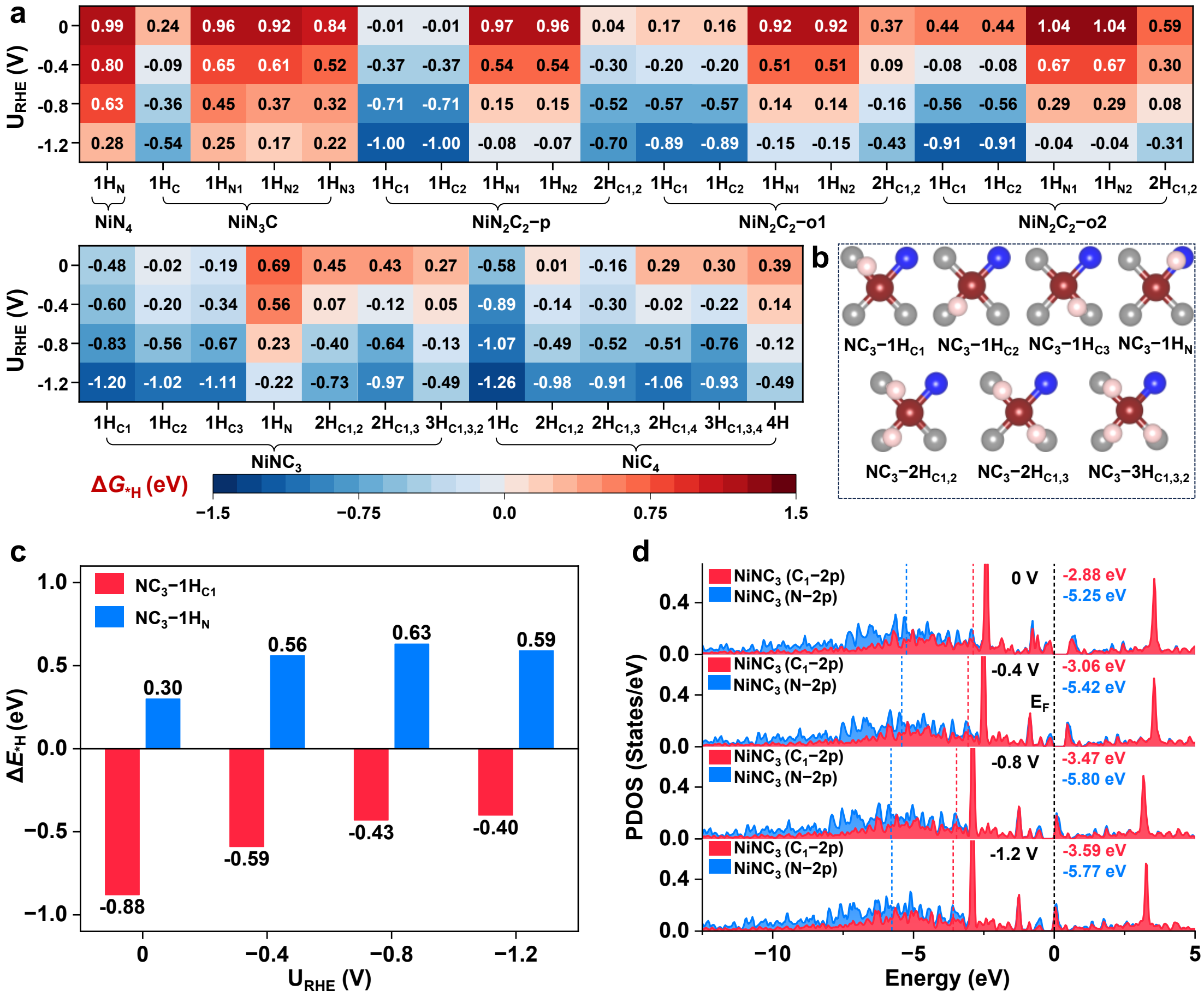

**Figure 2.** (a) Free energy change ($\Delta G_{*H}$) for hydrogen adsorption at C and N sites of $NiN_{4-x}C_x$ systems under various electrode potentials (0, −0.4, −0.8, and −1.2 $V_{RHE}$). (b) Representative atomic configurations of H adsorption at the C and N sites of $NiNC_3$, where Ni, H, N, and C atoms are shown in vermilion, pink, blue, and gray, respectively. (c) Adsorption energies ($\Delta E_{*H}$) of H at $C_1$ and N sites in $NiNC_3$ under various electrode potentials. (d) Projected density of states (PDOS) of the p orbitals of the $C_1$ and N sites under various electrode potentials (vs. RHE), with the p-band centers indicated by red and blue dashed lines for the $C_1$ and N sites, respectively.

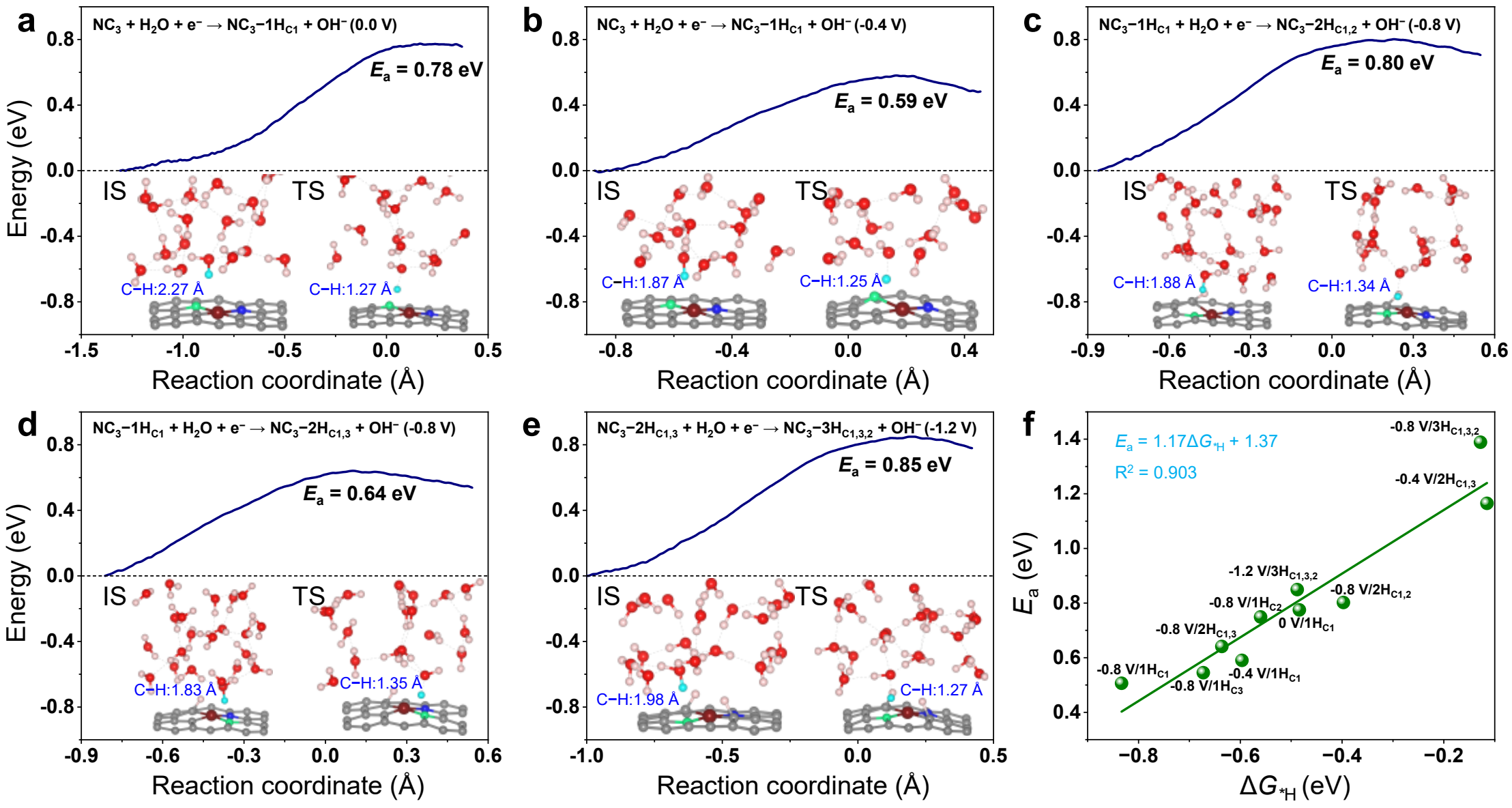

**Figure 3.** Slow-growth ab initio molecular dynamics simulations for the hydrogenation of the C sites of $NiNC_3$ under different electrode potentials. Free energy profile for the first hydrogenation of the $C_1$ site at 0 $V_{RHE}$ (a), $C_1$ site hydrogenation at −0.4 $V_{RHE}$ (b), the second hydrogenation at $C_2$ (c) and $C_3$ (d) sites at −0.8 $V_{RHE}$, and the third hydrogenation at the $C_2$ site at −1.2 $V_{RHE}$ (e). (f) The Brønsted-Evans-Polanyi (BEP) relation between hydrogenation barriers ($E_a$) and adsorption free energies ($\Delta G_{*H}$).

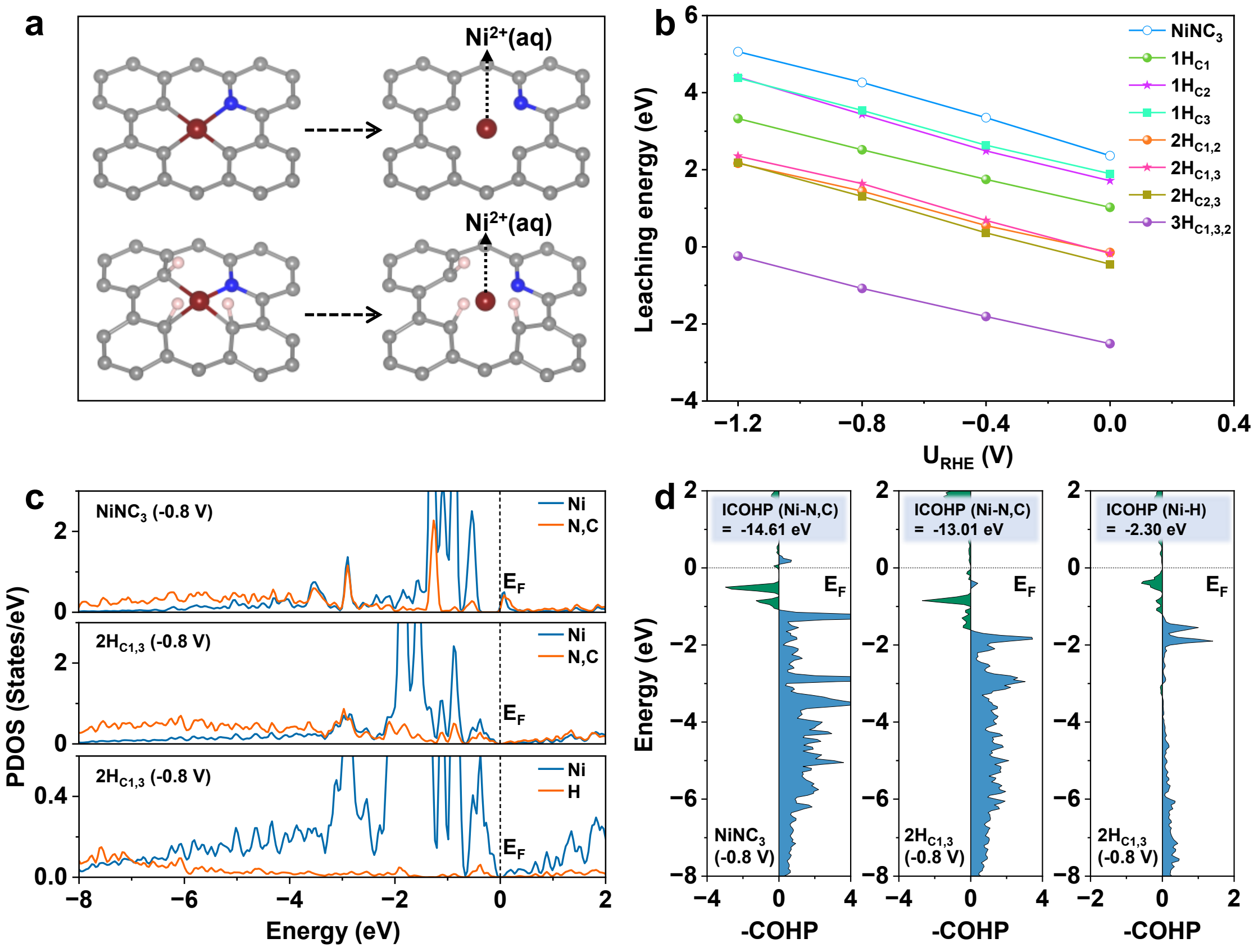

**Figure 4.** (a) Schematic illustration of the Ni leaching process from $NiNC_3H_y$: $NiNC_3H_y \leftrightarrow NC_3H_y + Ni^{2+}(aq) + 2e^-$. (b) Calculated leaching energies for $NiNC_3$ and its hydrogenation configurations under various electrode potentials. (c) Projected density of states (PDOS) of Ni, N, C, and H atoms for bare $NiNC_3$ and the hydrogenated configuration ($NiNC_3$-$2H_{1,3}$) at −0.8 $V_{RHE}$. (d) Crystal orbital Hamilton population (COHP) analysis of Ni−N,C and Ni−H interactions for the bare $NiNC_3$ and $NiNC_3$-$2H_{1,3}$ at −0.8 $V_{RHE}$, where positive and negative values denote bonding and antibonding contributions, respectively. In (c) and (d), the Fermi level is set to 0 eV.

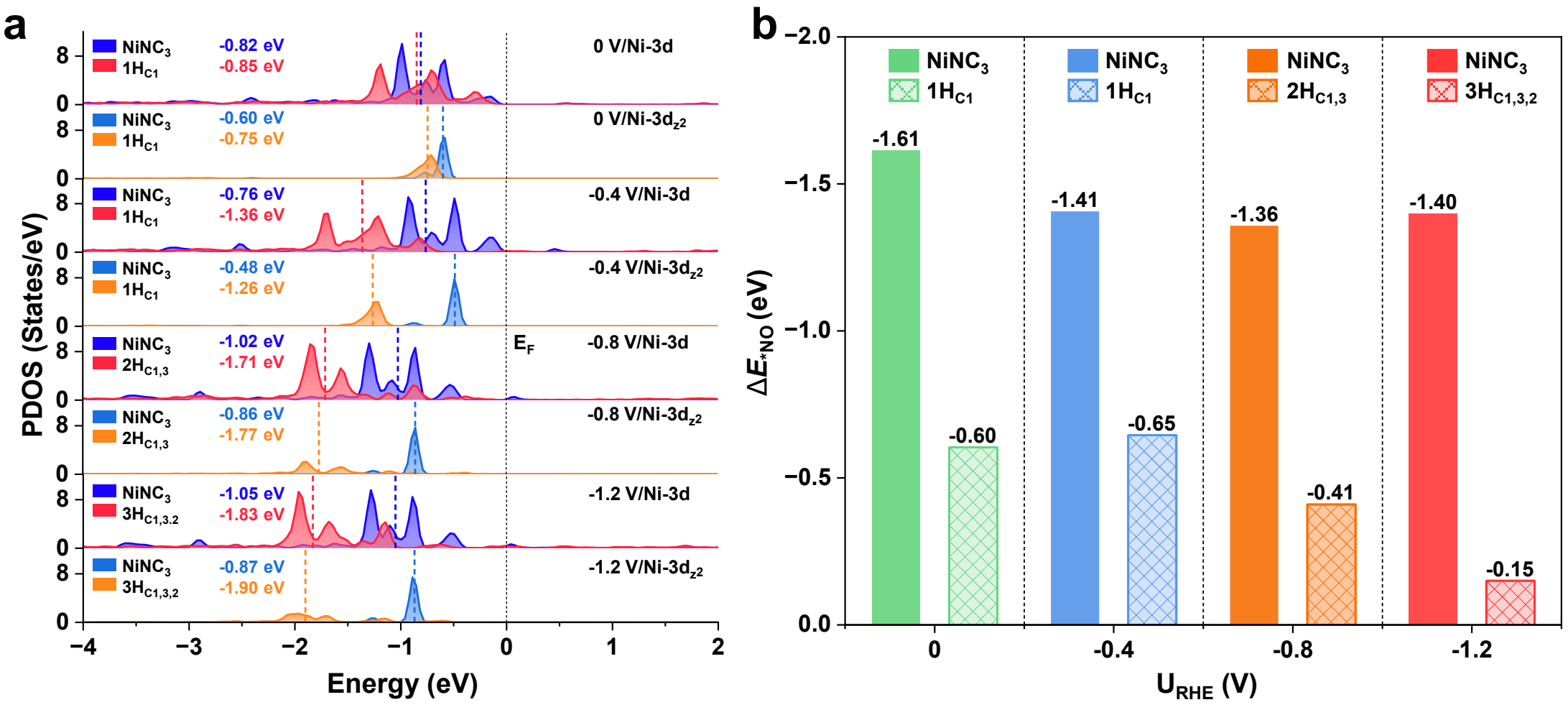

**Figure 5.** (a) Projected density of states (PDOS) of the Ni 3d and $d_{z^2}$ orbitals for the bare $NiC_3$ and the hydrogenated configurations ($NiNC_3$-1$HC_1$, $NiNC_3$-2$HC_{1,3}$, and $NiNC_3$-3$HC_{1,3,2}$) under electrode potentials of 0, −0.4, −0.8, and −1.2 $V_{RHE}$. The d-band and $d_{z^2}$-band centers are indicated by dashed lines. The Fermi level is set to 0 eV. (b) Calculated adsorption energy ($\Delta E_{*NO}$) for NO on the bare and hydrogenated $NiNC_3$ at specific electrode potentials.

**Potential-Induced Dynamic Coordination of Nonmetal Atoms Directly Bound to Metal Centers in Graphene-Embedded Single-Atom Catalysts and Its Implications**

Jiahang Li[1], Suhang Li[1], Chong Yan[1], Qinzhuang Liu[1], Jiajun Yu[1*], Dongwei Ma[1*]

[1]Anhui Provincial Collaborative Innovation Center for Advanced Functional Composite Materials, *College of Physics and Electronic Engineering, Huaibei Normal University, Huaibei 235000, China*

# Supporting Information

*Corresponding author. E-mail: yujiajun@chnu.edu.cn (J. Yu)
*Corresponding author. E-mail: madw@chnu.edu.cn, dwmachina@126.com (D. Ma)

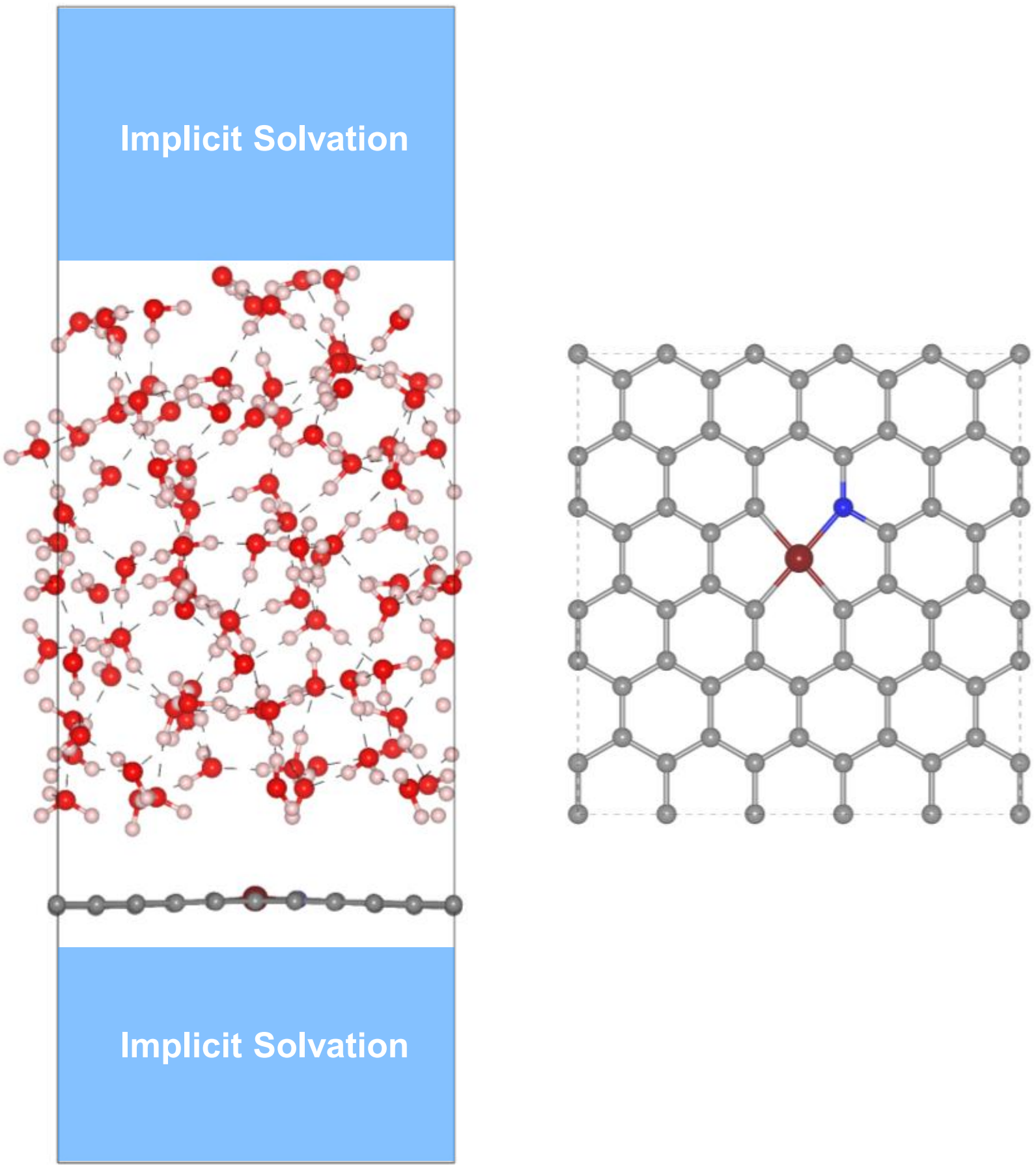

**Fig. S1.** Side (left) and top (right) views of the $NiNC_3$ system with 90 explicit water molecules. A random $H_2O$ configuration was first equilibrated by a 1 ns force-field MD simulation using LAMMPS,[1] providing the initial structure for the subsequent AIMD calculations. Hydrogen bonds are represented by the dashed lines. Color code: C, gray; N, blue; Ni, brown; O, red; H, pink. The supercell was constructed such that the confined water density matched the bulk value, with vacuum regions of ~8 Å introduced above the water layer and below the graphene surface to ensure a well-converged vacuum level at $z = 0$.

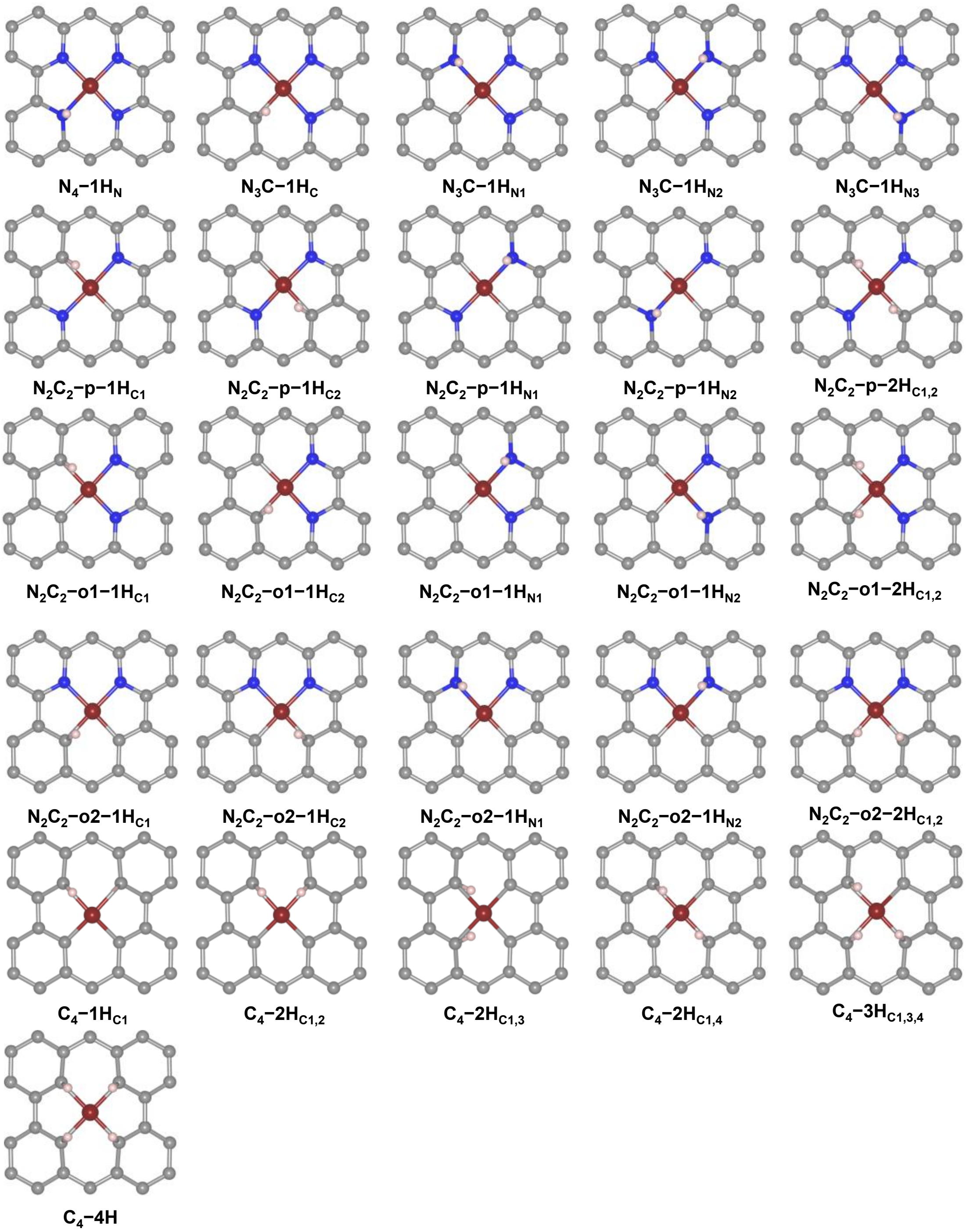

**Fig. S2.** Configurations of $NiN_xC_{4-x}$ with H adsorbed at the C and N sites.

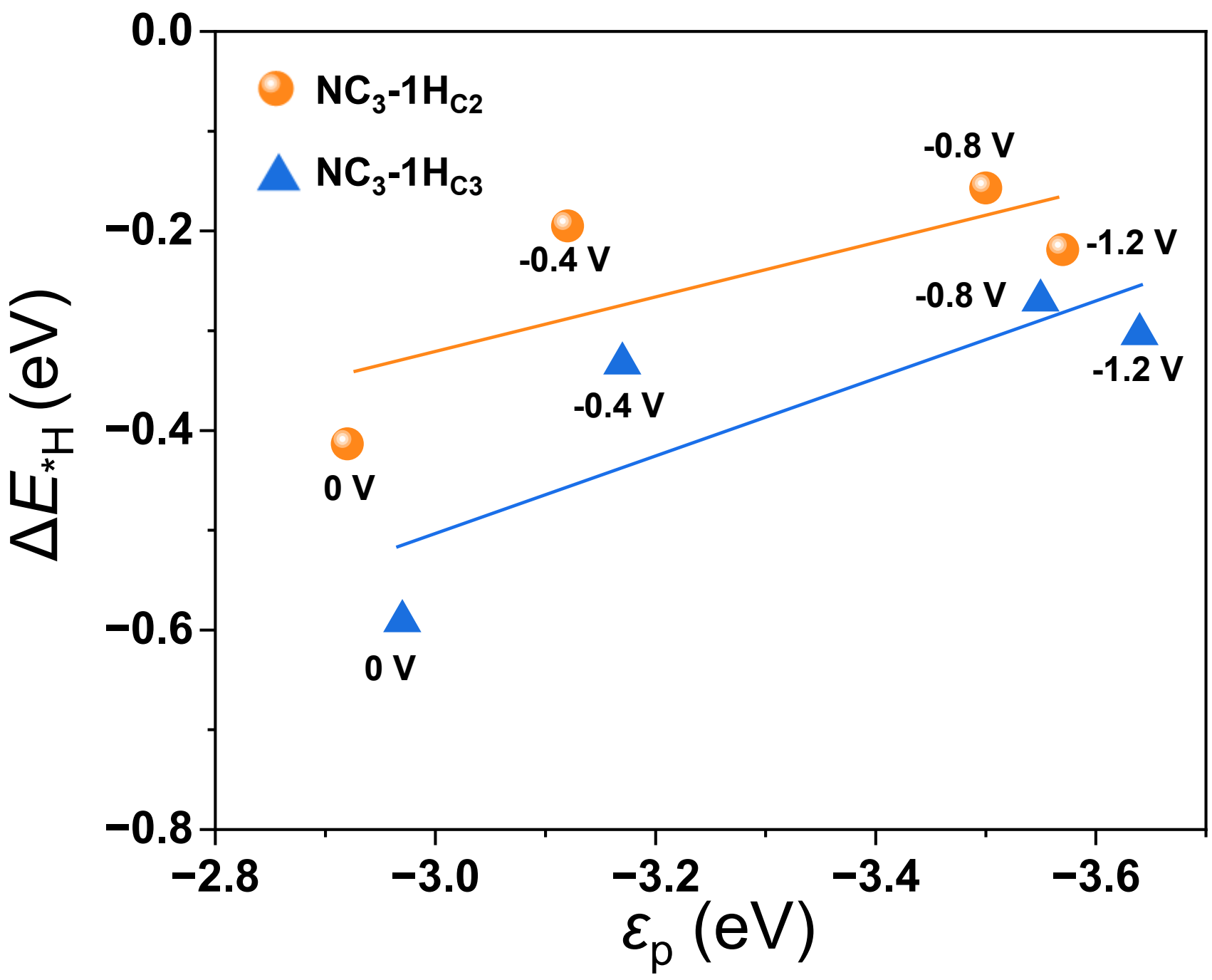

**Fig. S3.** Relationship between hydrogen adsorption energy and the p-band center of the C site in $NiNC_3$.

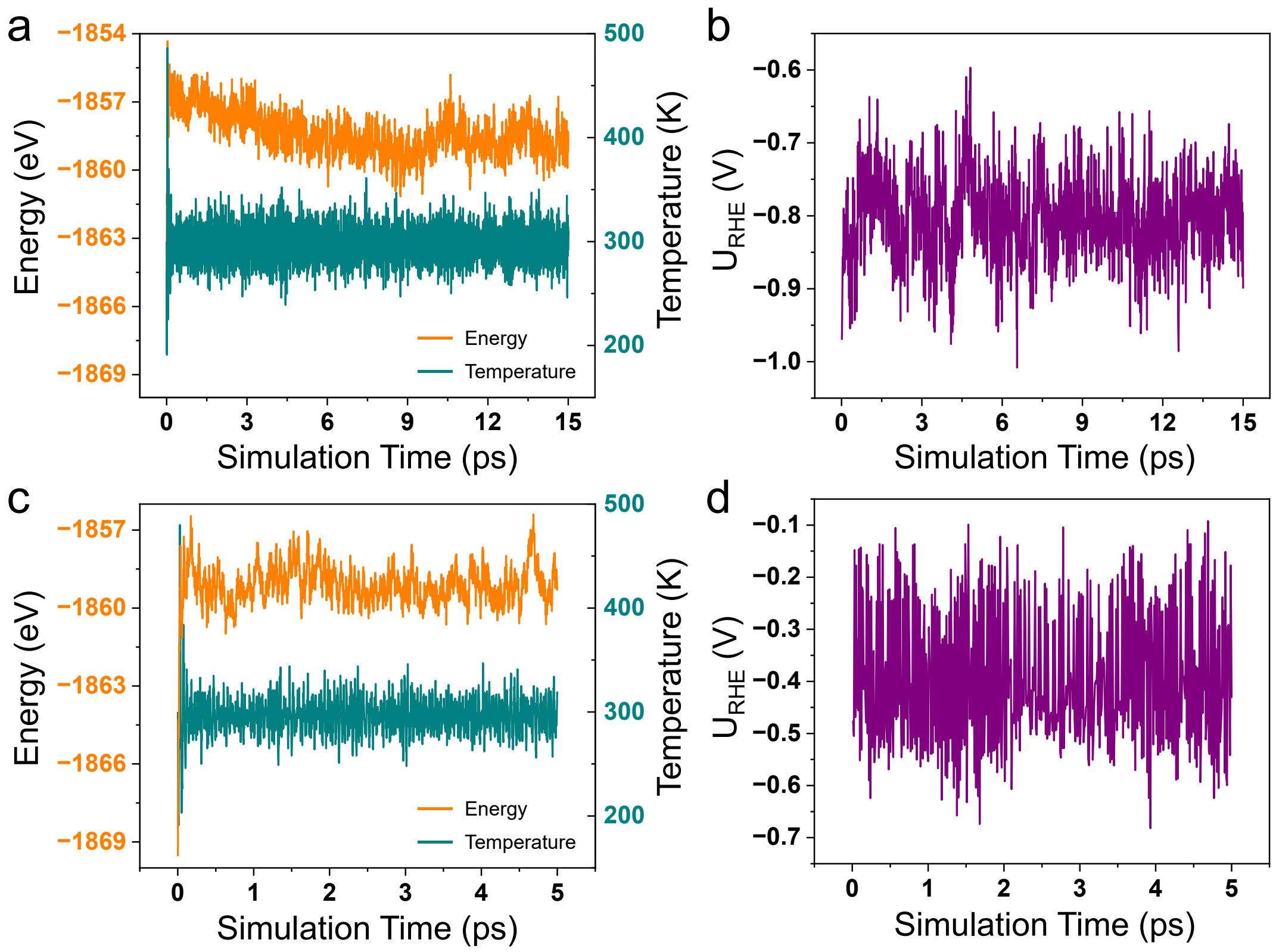

**Fig. S4.** Time evolution of energy and temperature (a) and electrode potential (b) along a 15 ps AIMD trajectory for $NiNC_3$ at $U_{RHE}$ = −0.8 V. Panels (c) and (d) show the corresponding results for a 5 ps trajectory at $U_{RHE}$ = −0.4 V.

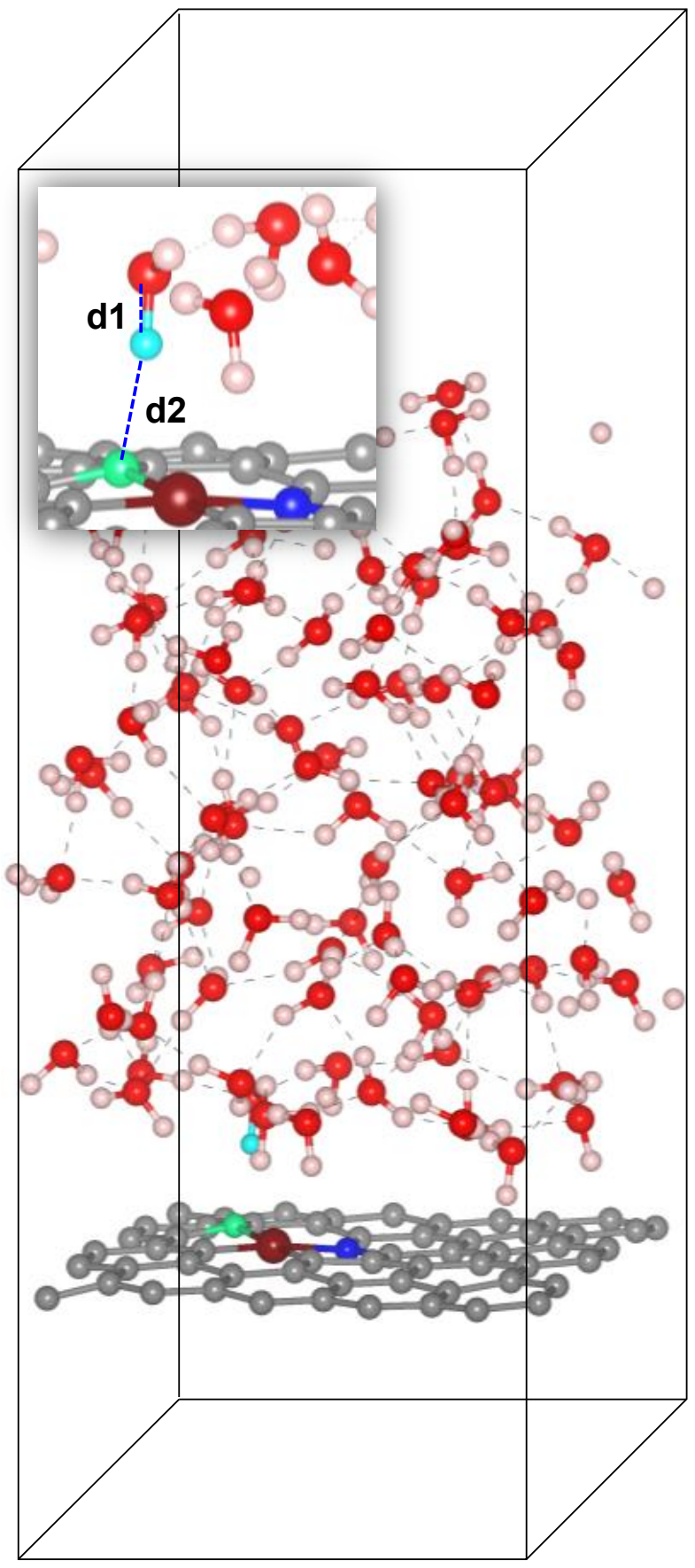

**H*and OH⁻ formation: CV = d1 – d2**

**Fig. S5.** Schematic of individual and collective variables used as reaction coordinates in the "slow-growth" approach for H* and $OH^-$ formation.

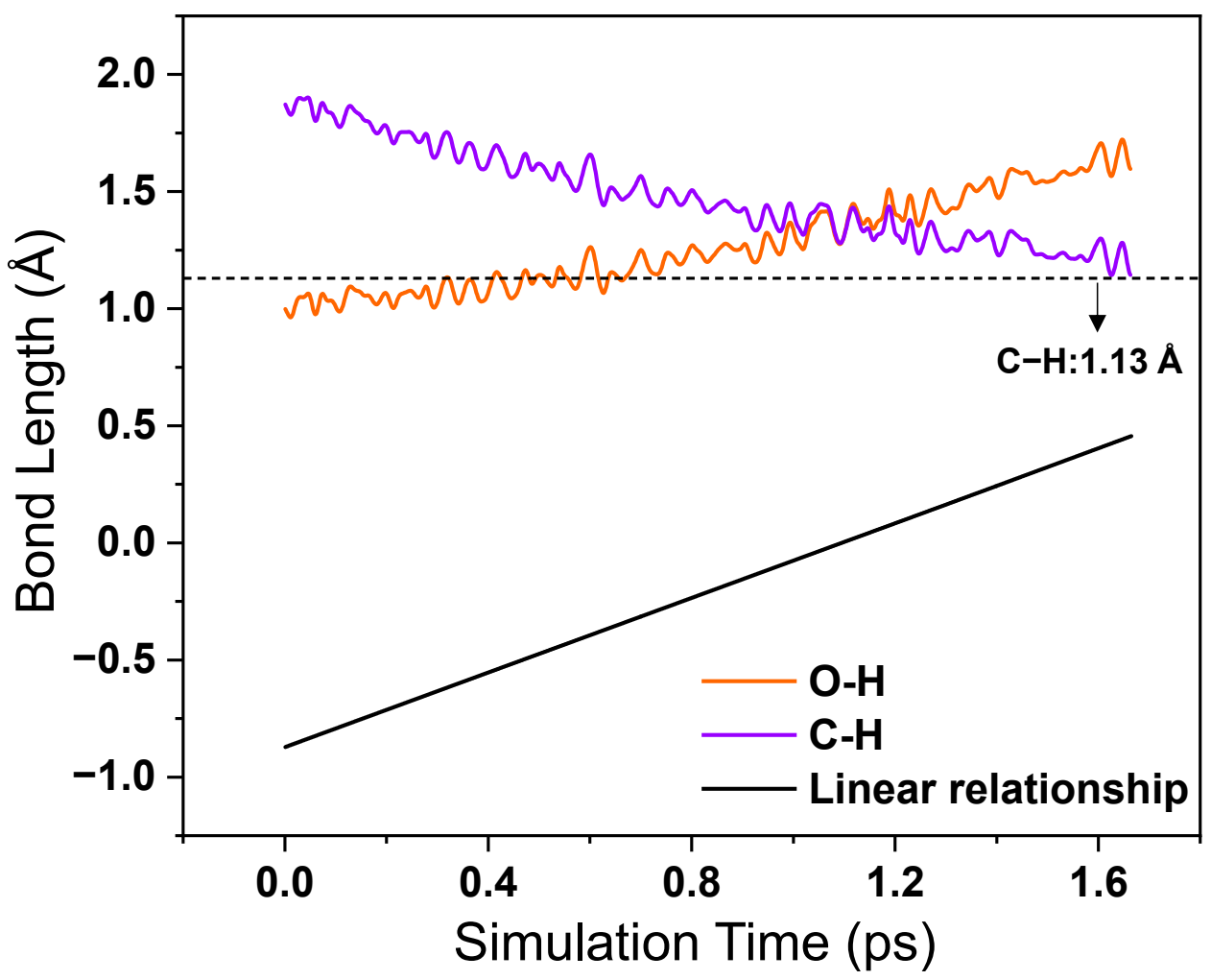

**Fig. S6.** The dynamic evolution of C–H and O–H bond lengths during the first H adsorption and $OH^-$ formation at the C1 site of the $NiNC_3$ under –0.4 $V_{RHE}$ through slow-growth simulations. The reaction coordinate follows a linear trend (black solid line), and the optimized C−H bond length (~1.13 Å) is indicated by the black dotted line.

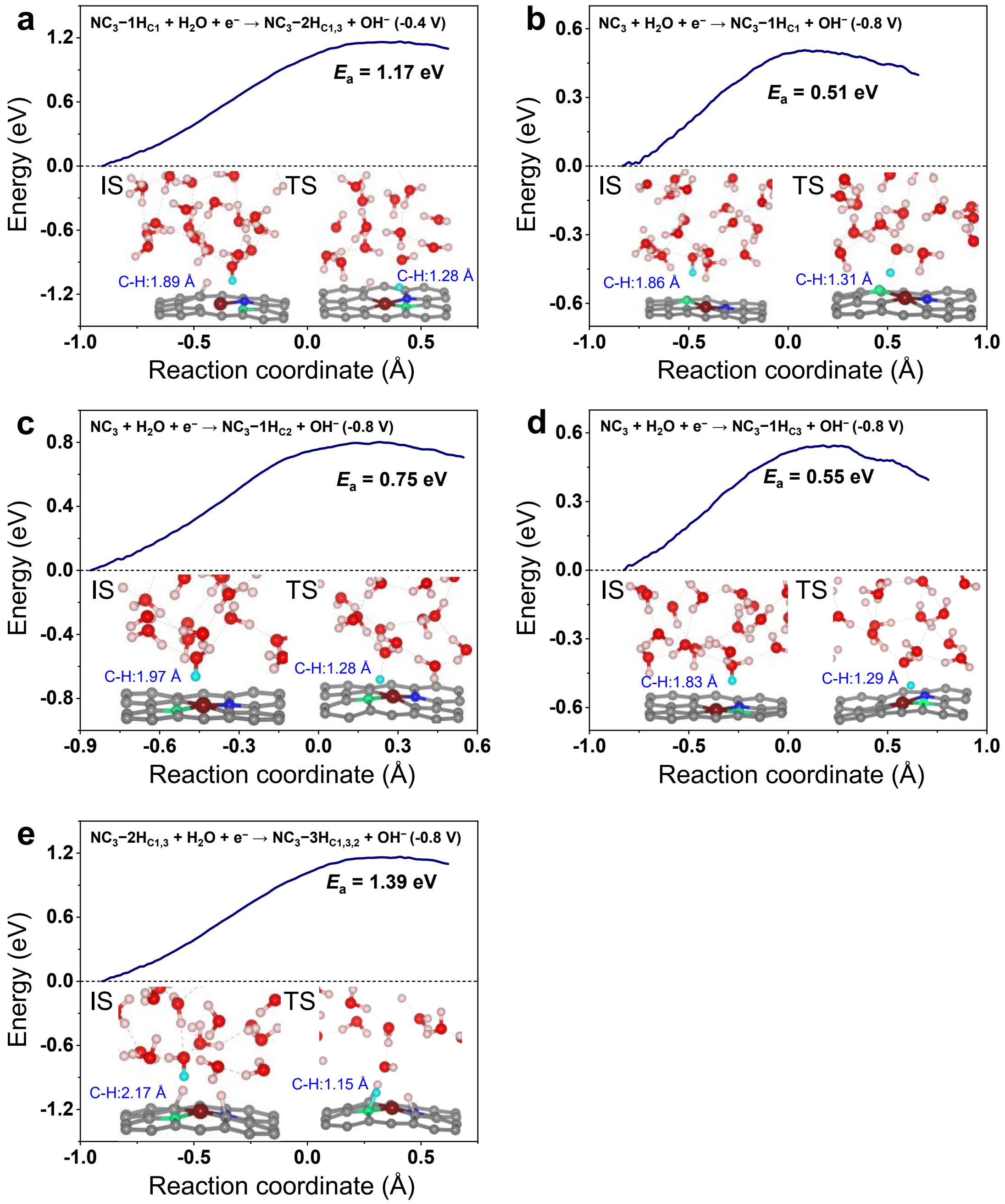

**Fig. S7.** Slow-growth ab initio molecular dynamics simulations of the hydrogenation at the C sites of $NiNC_3$ under different electrode potentials. Free energy profile for the first hydrogenation at the $C_1$ site at −0.4 $V_{RHE}$ (a), hydrogenation of the $C_1$, $C_2$, and $C_3$ sites at −0.8 $V_{RHE}$ (b–d), and the third hydrogenation at the $C_2$ site at −0.8 $V_{RHE}$ (e).

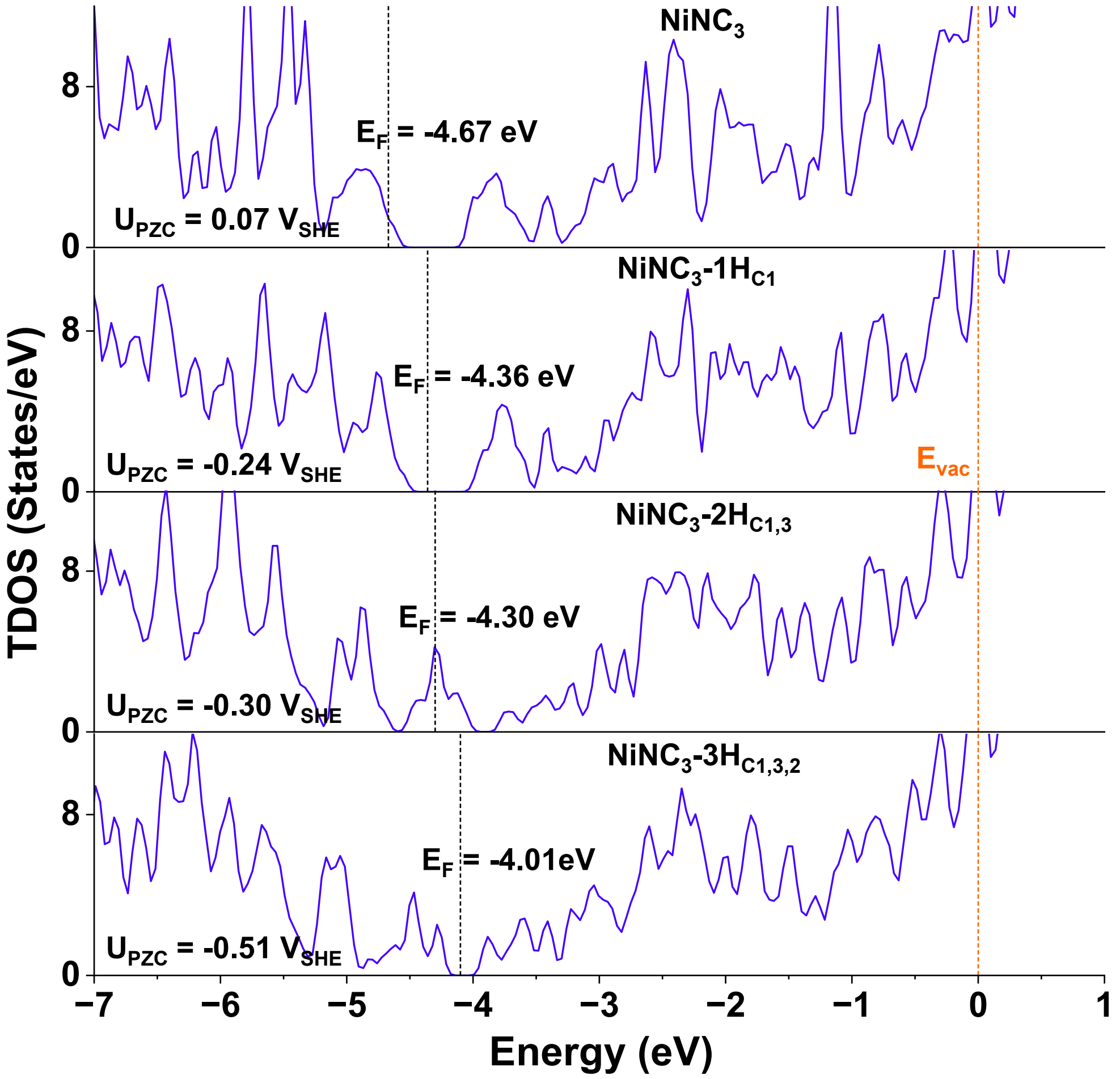

**Fig. S8.** Total density of states (TDOS) of the bare $NiNC_3$ system and its hydrogenated configurations. The vacuum energy level ($E_{VAC}$) is set to zero, and the Fermi level ($E_F$) is indicated by black dashed lines. The potential of zero charge ($U_{PZC}$) is also shown.

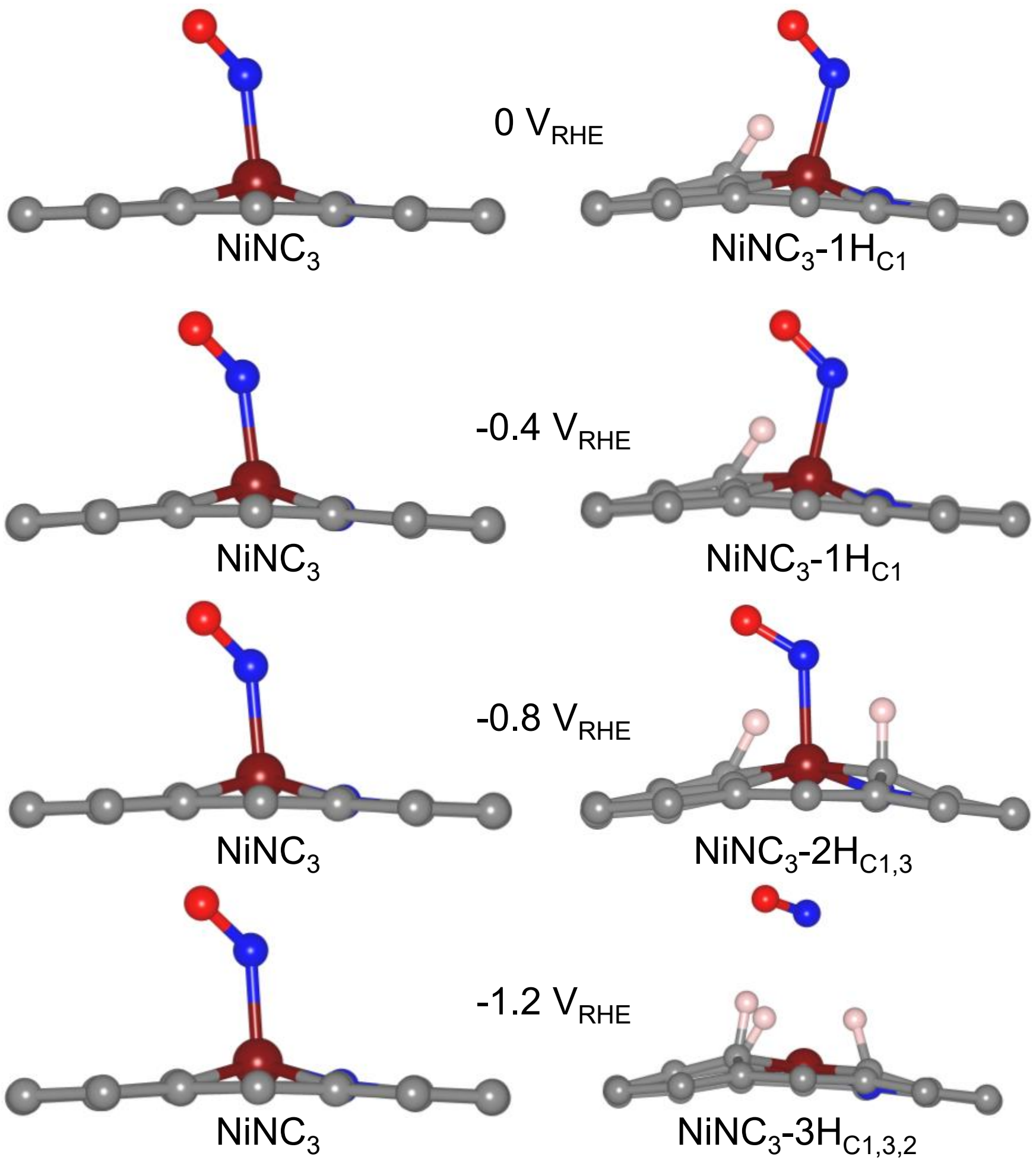

**Fig. S9.** NO adsorption configurations on $NiNC_3$, $NiNC_3$-$1H_{C1}$, $NiNC_3$-$2H_{C1,3}$, and $NiNC_3$-$3HC_{1,3,2}$ under specific electrode potentials.

**Table S1.** The calculated H adsorption free energy ($\Delta G_{*H}$) and corresponding energy barrier ($E_a$) for the Ni-coordinated C site in the $NiNC_3$ system under varying electrode potentials (0, –0.4, –0.8, and –1.2 $V_{RHE}$).

| $U_{RHE}$ | $\Delta G_{*H}$ (eV) | $E_a$ (eV) |
|---|---|---|
| 0 V/1$H_{C1}$ | -0.48 | 0.78 |
| -0.4 V/1$H_{C1}$ | -0.60 | 0.59 |
| -0.4 V/2$H_{C1,3}$ | -0.12 | 1.17 |
| -0.8 V/1$H_{C1}$ | -0.83 | 0.51 |
| -0.8 V/1$H_{C2}$ | -0.56 | 0.75 |
| -0.8 V/1$H_{C3}$ | -0.67 | 0.55 |
| -0.8 V/2$H_{C1,2}$ | -0.40 | 0.80 |
| -0.8 V/2$H_{C1,3}$ | -0.64 | 0.64 |
| -0.8 V/3$H_{C1,3,2}$ | -0.13 | 1.39 |
| -1.2 V/3$H_{C1,3,2}$ | -0.49 | 0.85 |

**Note 1.** The leaching process of Ni single-atom from the $NiN_{4-x}C_xH_y$ to form solvated $Ni^{2+}$(aq) can be expressed as

$$NiN_{4-x}C_xH_y \leftrightarrow N_{4-x}C_xH_y + Ni^{2+}(aq) + 2e^- \quad (a)$$

The free energy change for the above process can be calculated as

$$\Delta G = G(N_{4-x}C_xH_y) + G(Ni^{2+}(aq) + 2e^-) - G(NiN_{4-x}C_xH_y)$$

$$= G(N_{4-x}C_xH_y) + G(Ni^{2+}(aq)) + 2\mu_e - 2eU_{RHE} - G(NiN_{4-x}C_xH_y) \quad (1)$$

Similar to the computational hydrogen electrode model [2], the free energy of the Ni ion, $G(Ni^{2+}(aq)$, can be obtained from the experimental standard hydrogen electrode $U_0$ (0.257 V) and the calculated free energy of the bulk metal $G(Ni(s))$ as follows

$$Ni^{2+}(aq) + 2e^- \leftrightarrow Ni(s) \quad (b)$$

$$G(Ni(s)) - G(Ni^{2+}(aq)) - 2\mu_e + 2eU_0 = 0 \quad (2)$$

$$G(Ni^{2+}(aq)) + 2\mu_e = G(Ni(s)) + 2eU_0 \quad (3)$$

So, the corresponding free change at a constant potential $U_{RHE}$ can be calculated as

$$\Delta G = G(N_{4-x}C_xH_y) + G(Ni(s)) + 2eU_0 - 2eU_{RHE} - G(NiN_{4-x}C_xH_y) \quad (4)$$